\documentclass[fleqn,10pt]{olplainarticle}
 
\usepackage{float}

\title{GLOSSA: a user-friendly R Shiny application for Bayesian machine learning analysis of marine species distribution}

\author[1,2,*]{Jorge Mestre-Tomás}
\author[1,3]{Alba Fuster-Alonso}
\author[4]{José M. Bellido}
\author[1,5]{Marta Coll}

\affil[1]{Institute of Marine Sciences (ICM-CSIC), Renewable Marine Resources Department, Barcelona, Spain.}
\affil[2]{Universitat Politècnica de València(UPV), Department of Applied Statistics and Operational Research, and Quality, Valencia, Spain.}
\affil[3]{University of Valencia (UV), Department of Statistics and Operations Research (VaBar), Valencia, Spain.}
\affil[4]{Spanish Institute of Oceanography (IEO-CSIC), Murcia, Spain.}
\affil[5]{Ecopath International Initiative (EII), Barcelona, Spain.}
\affil[*]{Corresponding author: jorgemestre@icm.csic.es}


\begin{abstract}

\textbf{1.} Species distribution models (SDMs) are one of the most common statistical methods to assess species occupancy and geographic distribution patterns. With the increasing complexity of ecological data, many methodological approaches have been developed, often accessible through command-line interfaces or graphical user interfaces (GUIs). However, few species distribution modeling tools are designed to be well-documented, user-friendly, flexible, and reproducible.

\textbf{2.} Here we introduce GLOSSA, an open-source R package and Shiny app designed for species distribution modeling using species occurrence and environmental data. GLOSSA's user-friendly interface guides users through steps including data uploading, processing, model fitting, spatial and temporal projections, and interactive visualization of results. The app also calculates variable importance, generates response curves with environmental variables, and performs cross-validation. At its core, GLOSSA modeling approach is based on Bayesian Additive Regression Trees (BART), an innovative machine learning method.

\textbf{3.} We present the functionality and versatility of GLOSSA through three case studies, addressing a range of ecological scenarios at regional and global scales. Along with comprehensive documentation, examples, and tutorials, these case studies illustrate how an intuitive graphical interface can make species distribution modeling accessible to a broad audience.

\textbf{4.} GLOSSA stands out as an easy-to-use tool for species distribution modeling, providing an intuitive interface, detailed documentation, flexible modeling, and interactive result exploration and export options. Additionally, its outputs can be used directly to inform marine ecosystem models (MEMs), enhancing its utility in ecological research and applications.

\end{abstract}

\begin{document}

\flushbottom
\maketitle
\thispagestyle{empty}

\begin{center}
    \textbf{Note.} This is a preprint.
\end{center}


\section*{Introduction}

Understanding the geographic distribution of species over time and the effects of environmental conditions on them is crucial for ecological research, biodiversity studies, and conservation management \citep{pulliam2000relationship, guisan2013predicting, franklin2013species, srivastava2019species}. Species distribution models (SDMs) and ecological niche models (ENMs) are widely used to link species occurrence data with environmental variables and identify potentially suitable habitats where the species can survive. SDMs have a wide range of applications, including predicting climate change effects \citep{brown2016ecological, booth2018species, melo2020ecological}, assessing invasive species risks \citep{fournier2019predicting, lyons2020identifying}, understanding predator-prey dynamics \citep{liu2023development, ge2024robust, hill2024past}, and planning marine protected areas \citep{ferrari2018integrating, hunt2020identifying}. Beyond ecology, SDMs have also been applied in epidemiology \citep{gosoniu2006bayesian, martinez2018spatial} and paleontology \citep{svenning2011applications, eduardo2018extending}, among many other disciplines.

The increasing availability of georeferenced species and environmental data, together with advances in computational power, has led to an increase in the use of SDMs \citep{tyberghein2012bio, GBIF}. However, this amount of data also presents challenges, particularly in terms of the usability of SDMs by a wider audience who may lack advanced training in specific techniques \citep{wallace2012closing}. This gap highlights the need for accessible and user-friendly tools.

To meet the growing demand for easy-to-use methods, several tools have developed graphical user interfaces (GUIs) that simplify model use - see, for example, \cite{osorio2020ntbox}, \cite{kass2023wallace}, \cite{coro2024open} or \cite{figueira2024shiny}. Despite these advances; there is ongoing work to ensure comprehensive documentation, make the model and underlying process more transparent to users, and include novel methodologies \citep{mislan2016elevating}.

GLOSSA (GLOBAL SPECIES SPATIOTEMPORAL ANALYSIS) is an R Shiny application designed to make species distribution modeling more accessible and user-friendly, particularly in marine environments. GLOSSA allows users to fit SDMs using species occurrences and environmental data without requiring advanced coding skills. Wrapped in an R package \citep{Rcoreteam, shiny}, GLOSSA enables the analysis of past, present, and future species distributions by providing an intuitive and interactive GUI combined with comprehensive documentation and tutorials. The GLOSSA workflow involves: (1) loading georeferenced occurrences and environmental data, (2) filtering occurrences and processing data layers, (3) model fitting and prediction, and (4) interactive visualization and export of results.

GLOSSA modeling approach is based on Bayesian Additive Regression Trees (BART, \citealp{chipman2010bart}), a flexible, non-parametric sum-of-trees model that has been shown to be useful in modeling complex ecological relations at regional and global scales \citep{thompson2023climate, fuster2024machine}. The BART model uses a Bayesian framework to generate posterior predictive distributions for every pixel of the study area, allowing GLOSSA to provide detailed descriptions of species distributions, including uncertainty metrics.

In this paper, we present GLOSSA as a novel and easy-to-use Shiny application for SDMs that uses Bayesian machine learning models. First, we describe the GLOSSA workflow and implementation. Next, we provide detailed examples across three case studies that show its capabilities in species distribution modeling. Finally, we discuss the overall utility of the GLOSSA approach.

\section*{Overview of GLOSSA}

\subsection*{Workflow}

\begin{figure}[ht]
\centering
\includegraphics[width=\linewidth]{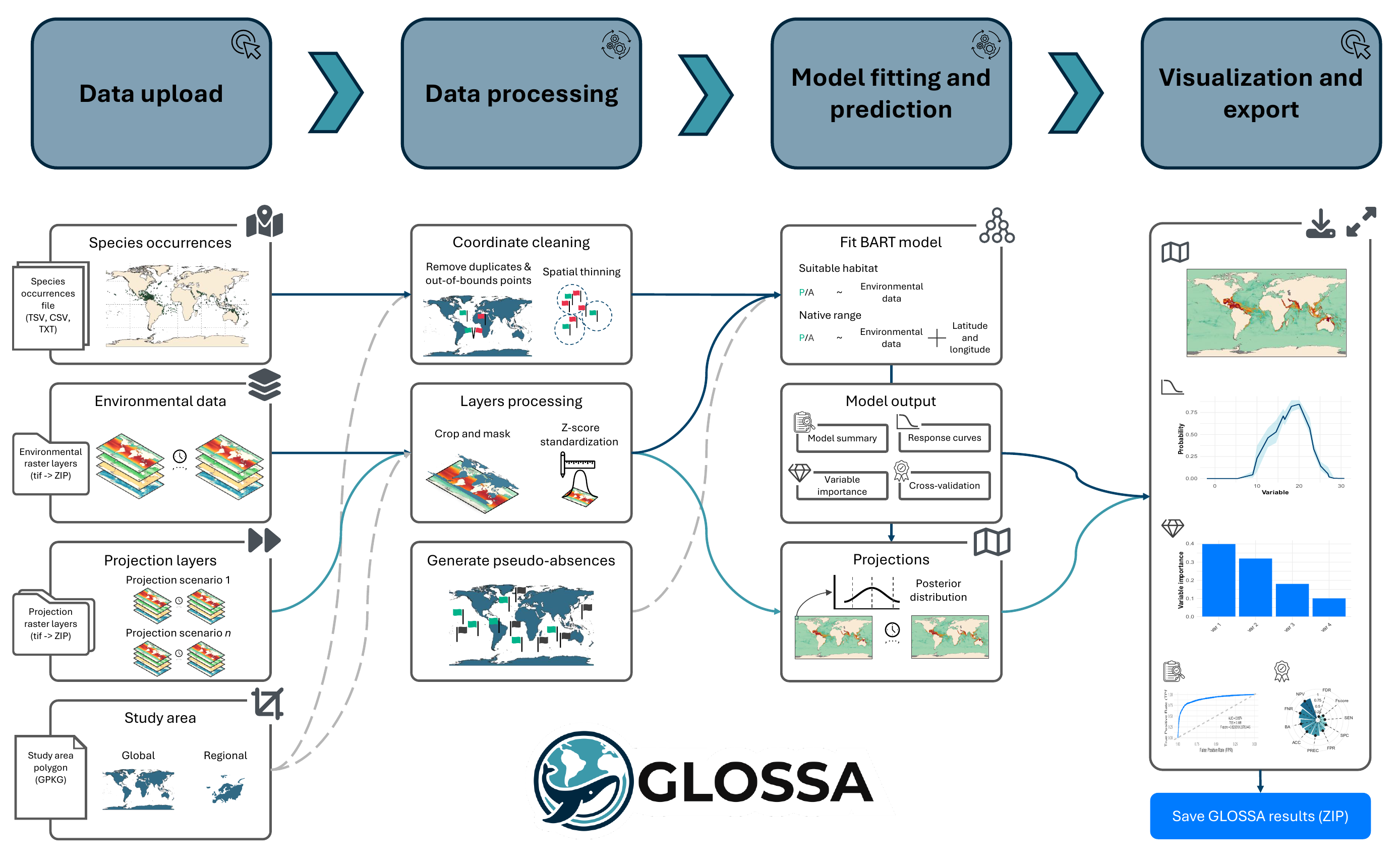}
\caption{Overview of the GLOSSA workflow.}
\label{fig:glossa_workflow}
\end{figure}

GLOSSA is designed to follow a linear analysis workflow (Figure \ref{fig:glossa_workflow}), supported by an intuitive user interface with a simple learning curve. The workflow includes four main steps: \textit{Data upload}, \textit{Data processing}, \textit{Model fitting and prediction}, and \textit{Visualization and export}. Users interact directly during the upload and visualization steps, while data processing and modeling are handled automatically. Below, we describe each step of the workflow

\begin{enumerate}
    \item \textit{Data upload}.

    \begin{enumerate}
        \item \textit{Species occurrences}. GLOSSA works with presence/absence data of species. Species occurrence data can come from different sources such as the Global Biodiversity Information Facility (GBIF, \url{https://www.gbif.org/}) or the Ocean Biogeographic Information System (OBIS, \url{https://obis.org/}). Though the modeling approach is based on presence-absence data, presence-only data can be provided, and balanced random pseudo-absence points will be generated across the study area. In addition, GLOSSA allows fitting independent models for multiple species in the same session. The GLOSSA website includes step-by-step tutorials on how to download data from global repositories and format the data to upload it to the app.
        
        \item \textit{Environmental data}. Environmental variables should be in raster format and can be sourced from databases such as the National Oceanic and Atmospheric Administration (NOAA, \url{https://www.noaa.gov/}), the Inter-Sectoral Impact Model Intercomparison Project (ISIMIP, \url{https://www.isimip.org/}), or Bio-ORACLE (\url{https://www.bio-oracle.org/}). GLOSSA extracts the values of these variables from each grid cell corresponding to the occurrence points, matching them according to the associated timestamps of the occurrence data if multiple time layers are provided. These layers are used to build the model matrix and also to define the study area unless a custom polygon is provided by the user. The analysis is performed only in areas where the environmental layers have complete data, avoiding regions with missing values.
        
        \item \textit{Projection layers}. GLOSSA allows for spatio-temporal predictions across the entire study area using the fitted model. Users can upload a series of projection layers representing different time periods for each environmental variable. GLOSSA will then generate predictions for each individual time step. Additionally, users can provide multiple time series representing different climate scenarios, and GLOSSA will predict species distributions for each scenario independently. This feature enables comprehensive forecasting under varying environmental conditions and future climate scenarios.
        
        \item \textit{Study area}. The user can provide a polygon to define the extent of the study area. This polygon will be used to filter species occurrences and to crop and mask environmental layers. If the polygon is not provided the analysis is performed where environmental data is available.
    \end{enumerate}
    
    \item \textit{Data processing}.

    \begin{enumerate}
        \item \textit{Coordinate cleaning}. After data input, GLOSSA automatically processes species occurrence coordinates by removing duplicates, records with missing coordinates, and points outside the defined study area, either outside the polygon boundaries or where environmental variables have missing values. Moreover, users can apply spatial thinning in specific cases to reduce the effects of sampling bias in clustered occurrences \citep{varela2014environmental}. This thinning process uses a precision-based approach from the GeoThinneR R package \citep{mestre2024geothinner}, allowing users to specify a precision level by defining the number of decimal digits to which coordinates are rounded for filtering.
        
        \item \textit{Layers processing}. Environmental layers are cropped and masked to match the study area polygon. Additionally, Z-score standardization can be applied, adjusting each variable to have a mean of 0 and a standard deviation of 1, when variables are on different scales.
        
        \item \textit{Generate pseudo-absences}. For presence-only data, GLOSSA generates pseudo-absences equal to the number of presences and randomly distributed across the study area. If occurrence data includes different timestamps, GLOSSA will generate pseudo-absences matching the number of presences for each specific times step.
    \end{enumerate}

    \item \textit{Model fitting and prediction}.

    \begin{enumerate}
        \item \textit{Fit BART model}. GLOSSA uses Bayesian Additive Regression Trees (BART, \citealp{chipman2010bart}) to fit species distribution models. Two types of models can be generated: (i) a suitable habitat model, based only on environmental data, and (ii) a native range model, which also includes spatial smoothing using latitude and longitude coordinates of historical data.
        
        \item \textit{Model output}. After fitting the model, GLOSSA determines an optimal classification cutoff using Youden's index \citep{youden1950index}, also known as true skill statistic (TSS), to predict potential presences and absences of the species. Then, a model summary is generated that includes a ROC curve with the area under the curve (AUC), TSS, and F-score metrics, the confusion matrix, and the distribution of the fitted values to illustrate the performance of the binary classifier model. If requested, the functional responses (response curves, the relationship between the occurrence of a species and each environmental variable) are computed as partial dependence plots \citep{friedman2001greedy}. These plots show the marginal effect that each variable has on the predicted outcome of the BART model. The response curves are calculated for the suitable habitat model without variable standardization so they can be interpreted on the original scale. In addition, variable importance is computed using a permutation-based approach \citep{wei2015variable}, measuring the change in the prediction error of the model using the F-score after we permuted the variable's values. Finally, a k-fold cross-validation with 5 folds is performed to evaluate the predictive capacity of the model.
        
        \item \textit{Projections}. When fitting a model, GLOSSA makes a single prediction of suitable habitat and/or native range models using an averaged environmental scenario. This scenario is created by calculating the mean values of each environmental variable across all provided time steps. This approach provides a general prediction rather than separate predictions for each time step. Additionally, projections can be made to different areas, time periods, and climate scenarios if projection layers are uploaded. These are represented with summary metrics of the posterior predictive distribution, taking advantage of the Bayesian framework.
    \end{enumerate}

    \item \textit{Visualization and export}. After completing the analysis, users can explore the results through interactive visualizations. These visualizations allow for an in-depth examination of the species distribution models and their projections. Users can easily export the results for further analysis or reporting.
\end{enumerate}

\subsection*{Modeling approach}

Currently, the approach used by GLOSSA for species distribution modeling is based on Bayesian Additive Regression Trees (BART; \citealt{chipman2010bart}), a promising approach in the field of machine learning for fitting SDMs. BART has begun to be widely used for SDMs, especially at a regional scale \citep{thompson2023climate}, and recently, it has been shown to perform well at a global scale \citep{fuster2024machine}. \citet{chipman2010bart} introduced BART as a non-parametric approach to regression modeling based on a sum-of-trees model where priors are used to regularize inference. The main advantage of BART is its ability to mitigate bias in predictions caused by the overfitting of regression trees. By using regularization priors, BART ensures each tree acts as a "weak learner", contributing only a small part of the overall model \citep{chipman2010bart}. This makes BART particularly suitable for species distribution modeling, which often involves complex, high-dimensional ecological data.

Since GLOSSA is based on presence and absence data, we use a logistic regression with a Bernoulli distribution to model the probability of presence. Therefore, the model applied is as follows:
\begin{equation}\label{eq:bart_model}
\begin{split}
 Y_{i} \sim Ber(\pi_{i}), \quad i = 1,...,n, \\
        \phi^{-1}(\pi_{i}) = \sum^{m}_{j}g_{j}(\textbf{X};T_{j},M_{j}), \hspace*{\fill}
\end{split}
\end{equation}
where $Y_{i}$ represents the presence/(pseudo-)absence of species for observation $i$, and $\pi_{i}$ is the parameter of interest representing the probability of presence linked to the predictor by a link function $\phi^{-1}$ that denotes the standard normal cdf (probit link function). Each $g_{j}$ corresponds to the $j-$th ($j=1,\ldots,m$) tree of the form $g_{j}(\textbf{X};T_{j},M_{j})$, where $m$ is the total number of trees, $\textbf{X}$ is a vector of multiple covariates, $T_{j}$ is a binary tree structure consisting of a set of interior node decision rules and a set of terminal nodes, and $M_{j} = \{\mu_{j},...,\mu_{jb}\}$ represents a set of parameter values associated with each of the $b_{j}$ terminal nodes of $T_{j}$. GLOSSA currently uses the default priors recommended by \citep{chipman2010bart} to regularize the fit. Finally, an MCMC algorithm is used to sample from the posterior distribution.

Regarding the model structure, we differentiate between two models in GLOSSA based on the predictor variables. When the model includes coordinates (i.e., longitude and latitude) as covariates, we refer to it as the "native range" model, which represents where the species is most likely to be found considering where it was found already. On the other hand, if the model is based only on key drivers such as temperature, salinity, or bathymetry, we refer to it as the "suitable habitat" model, which explains potential locations where conditions are appropriate for the target species, even if the species may not currently be present there for other reasons.

\subsection*{App features}

GLOSSA is implemented as an R package named \textit{glossa}, accessible on the Comprehensive R Archive Network (CRAN) at \url{https://cran.r-project.org/package=glossa} and on GitHub at \url{https://github.com/iMARES-group/glossa}. Its intuitive user interface is built using \textit{shiny} \citep{shiny}, \textit{bs4Dash} \citep{bs4Dash}, and \textit{shinyWidgets} \citep{shinyWidgets}, while the modeling approach relies on the \textit{dbarts} package \citep{dbarts}.

\begin{figure}[ht]
\centering
\includegraphics[width=\linewidth]{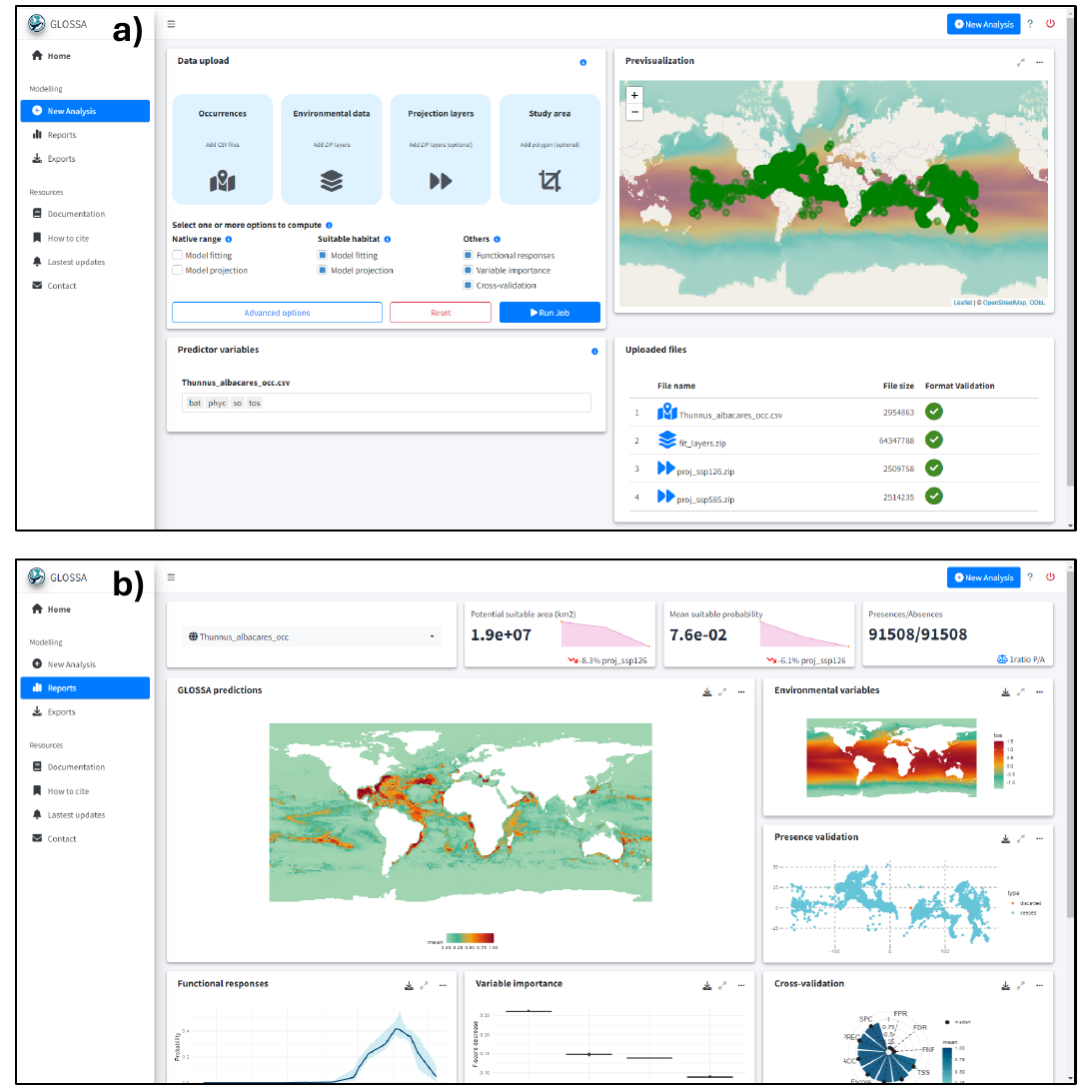}
\caption{Screenshots of GLOSSA’s user interface. (a) The \textit{New Analysis} window, where users can upload data, select analysis settings, and validate inputs. (b) The \textit{Reports} tab, displaying interactive visualizations of model results, including habitat suitability predictions and performance metrics.}
\label{fig:glossa_gui}
\end{figure}

The \textit{Home} tab of GLOSSA allows the user to access tutorial examples, start new analyses, and view detailed documentation. The quick guide provides an immediate reference for first-time users, while comprehensive documentation is available at \url{https://imares-group.github.io/glossa/}, offering in-depth documentation, tutorials, and case studies.

From the \textit{New Analysis} window (Figure \ref{fig:glossa_gui}a), users can upload data, select the analysis options, and pre-visualize the loaded data, including species occurrences, environmental variables, and the study area boundaries. A data validation table helps ensure that all inputs meet the required formats for analysis. Additionally, users can define different predictor variables for each species within the same session, providing flexibility in modeling more than one species dataset.

Once the analysis is completed, the \textit{Reports} tab (Figure \ref{fig:glossa_gui}b) provides interactive visualizations of suitable habitat and native range predictions across different time periods or climate scenarios, as well as plots of environmental variables and species occurrences. Users can explore different summary metrics from the posterior predictive distributions (mean, median, and quantiles), analyze trends in habitat suitability, and examine spatial and temporal patterns. Furthermore, GLOSSA provides model performance metrics, including ROC curves, the distribution of fitted values, and the classification of predicted outcomes. Functional responses are represented using the mean response curve, with a shaded area representing the 95\% credible interval. Variable importance is illustrated in a boxplot, summarizing the results of 10 iterations of permutation variable importance, sorted according to the mean importance. Additionally, k-fold cross-validation results are presented in a radar chart. All plot options can be accessed via the three-dot button in the top-right corner of each plot. Also, users are able to maximize plots for better visualization and export them in various formats.

The \textit{Export} window allows users to save all GLOSSA results, including data used to fit the models, raster predictions, functional responses, variable importance, and model validation metrics. These outputs are provided in standardized and interoperable formats to ensure compatibility with other downstream applications such as Marine Ecosystem Models (MEMs).

\section*{Worked examples}

We demonstrate the use of GLOSSA through three case studies involving marine species at global, regional, and local scales. Complete details for these examples, including data and code, are available in the vignettes provided in Supplementary Files S1–S3 and on the package website (\url{https://imares-group.github.io/glossa/}). The associated datasets can also be accessed via Zenodo (\url{https://doi.org/10.5281/zenodo.13833556}). Below, we provide a brief overview of each case study.

\subsection*{Case study 1. Climate change impacts on yellowfin tuna global distribution}

In our first example, we modeled the global habitat suitability of yellowfin tuna (\textit{Thunnus albacares}) under two different climate scenarios: SSP1-2.6, representing a sustainable future with reduced $\text{CO}_2$ emissions, and SSP5-8.5, a high-emission scenario characterized by increased global warming. Yellowfin tuna is an important migratory predator in the world's oceans and accounts for a significant portion of the global catch \citep{fao2024}, making it essential to understand how climate change could impact its large-scale distribution \citep{erauskin2019large}.

Occurrence data were obtained from the Ocean Biodiversity Information System (OBIS) \citep{OBIS_thunnus}, while environmental variables, including sea surface temperature ($^{\circ}$C), sea surface salinity (psu), primary productivity ($mmol/m^3$), and bathymetry (m), were obtained from the ISIMIP3b climate dataset using the GFDL-ESM4 Earth System Model \citep{warszawski2014inter}. Bathymetry data were downloaded from the ETOPO 2022 Global Relief Model developed by the National Oceanic and Atmospheric Administration (NOAA, \url{https://www.ncei.noaa.gov/products/etopo-global-relief-model}). Before fitting the model, a z-score standardization was applied to the variables.

\begin{figure}[H]
\centering
\includegraphics[width=\linewidth]{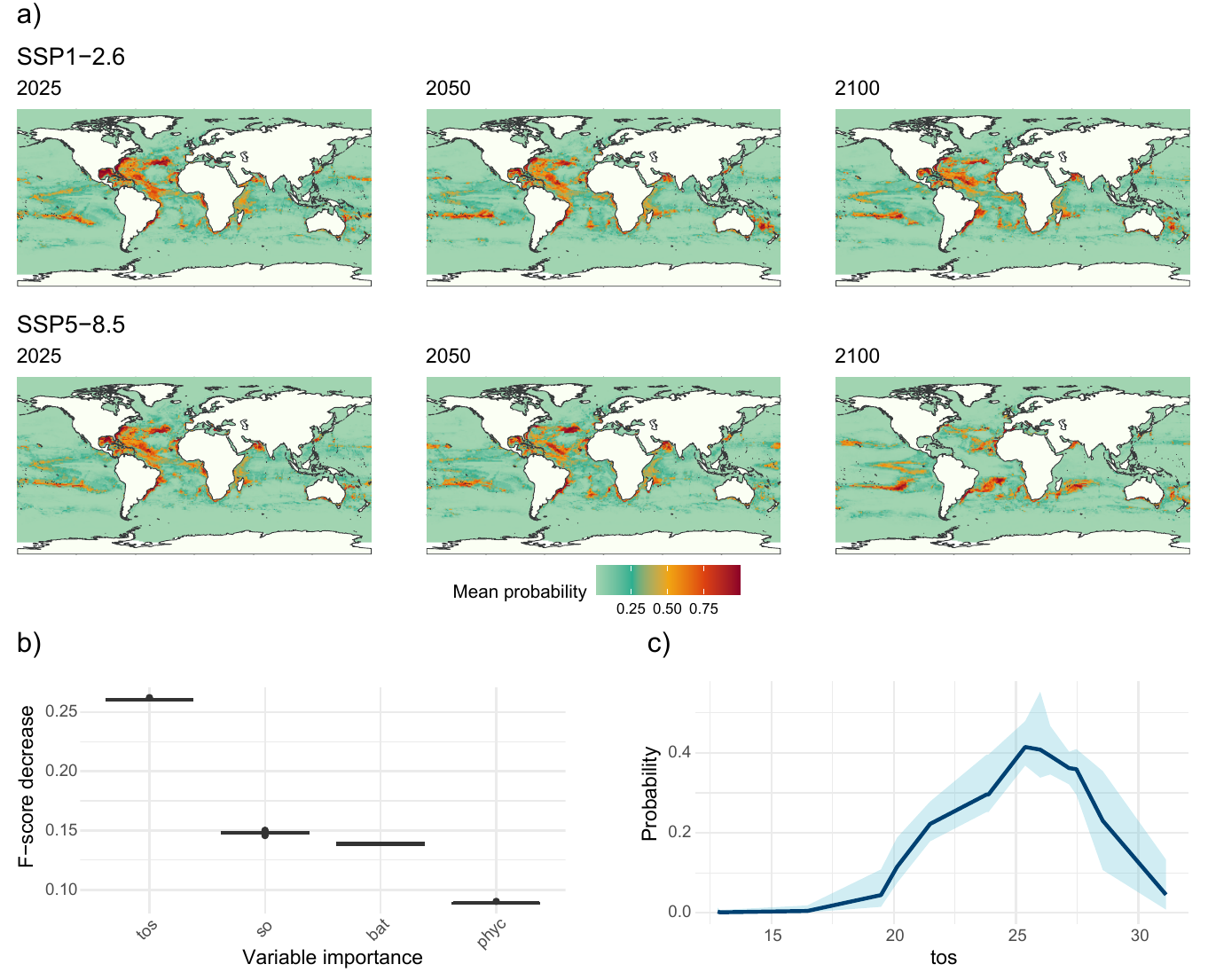}
\caption{a) Mean posterior probability of projected suitable habitats for \textit{Thunnus albacares} under the SSP1-2.6 and SSP5-8.5 climate scenarios on a global scale. b) Variable importance obtained from a permutation-based approach using the F-score (tos: sea surface temperature, so: salinity, bat: bathymetry, phyc: primary productivity). c) Response curve (partial dependence plot) for sea surface temperature.}
\label{fig:thunnus_albacares}
\end{figure}

The results show a decline in suitable habitat for \textit{Thunnus albacares} from 2025 to 2100 under both scenarios, with an 8.3\% reduction under SSP1-2.6 and a more intense 20\% reduction under SSP5-8.5 (Figure \ref{fig:thunnus_albacares}). Under the high-emission scenario, suitable habitats shift from equatorial and tropical regions toward the subtropics, reflecting the impact of increased warming on the distributional shift of the species.

Variable importance analysis, measured through permutation accuracy using the F-score, shows that sea surface temperature is the variable with the most important role in predicting species occurrences, followed by salinity and bathymetry, while primary productivity is less influential. The response curve for temperature indicates that the species' occurrence probability peaks around 25°C.

These findings highlight the role of temperature in shaping species distribution and the potential shifts and losses of suitable habitats under future climate scenarios of a widely distributed species. Additional results are provided in Supplementary File S1.

\subsection*{Case study 2. Suitable habitat of \textit{Caretta caretta} in the Mediterranean Sea}

This case study illustrates the application of GLOSSA within a user-defined region, focusing on the loggerhead turtle (\textit{Caretta caretta}) in the Mediterranean Sea. While globally listed as vulnerable by the IUCN Red List, the species is classified as least concern in the Mediterranean. However, climate change and global warming represent a risk on their life cycle and ecological niche \citep{mancino2022going}. Here, we analyze how different future climate scenarios will reshape the habitat suitability of the Mediterranean Sea for \textit{Caretta caretta}. We used occurrence data from GBIF \citep{GBIF_caretta}, and we filtered sighting records from 2000 to 2020. These data were combined with environmental variables from Bio-ORACLE v3.0 (\citealp{assis2024bio}, \url{https://www.bio-oracle.org/}) from the 2000-2010 and 2010-2020 decades. The environmental variables used in the analysis included sea surface temperature ($^{\circ}$C), primary productivity ($mmol/m^3$), salinity (psu), and bathymetry (m) from the ETOPO 2022 Global Relief Model by NOAA. After fitting the SDM with the standardized variables, we projected the suitable habitat under two different climate scenarios: SSP1-2.6 (sustainable development) and SSP5-8.5 (high emissions).

\begin{figure}[H]
\centering
\includegraphics[width=\linewidth]{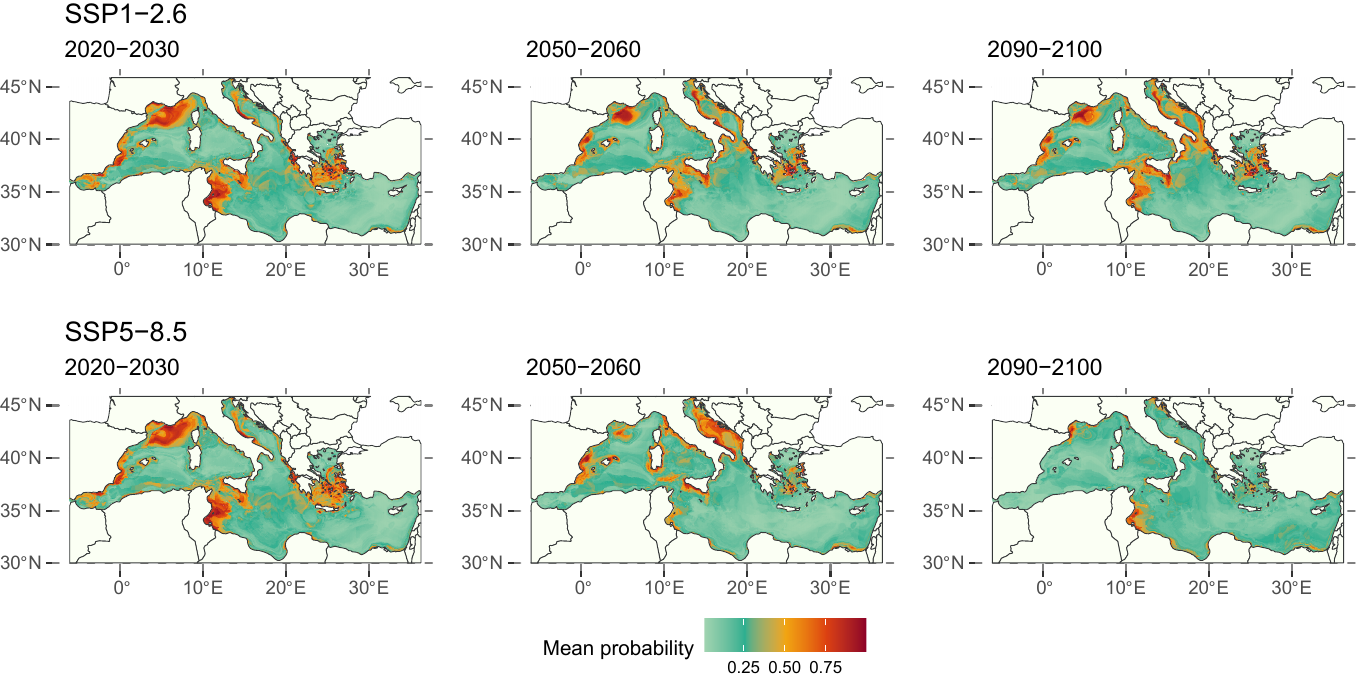}
\caption{Mean posterior probability of projected suitable habitats for \textit{Caretta caretta} under the SSP1-2.6 and SSP5-8.5 climate scenarios in the Mediterranean Sea.}
\label{fig:caretta_caretta}
\end{figure}

The results show a loss of habitat suitability in both scenarios from the 2020s to the 2090s (Figure \ref{fig:caretta_caretta}). However, the decline is much more pronounced under SSP5-8.5, where up to 80\% of suitable habitat area is lost. Additionally, habitat suitability in both scenarios tends to be higher in the western Mediterranean and the Adriatic Sea than in the eastern basin, where the loss of habitat is clear. Further details on the analysis and results can be found in Supplementary File S2.

\subsection*{Case study 3. Potential risk areas of \textit{Siganus luridus} in the Greek Seas}

Our final case study demonstrates GLOSSA’s application at a local scale, assessing potential invasion risks of the rabbit fish \textit{Siganus luridus} in the Greek Seas. This species is one of many Lessepsian migrants entering the Mediterranean through the Suez Canal, posing challenges to native species and making it crucial to predict their potential spread to inform management strategies \citep{solanou2023looking}. Occurrence data from 2000 to 2020 was obtained from an individual dataset from OBIS \citep{Sini2024}. The environmental variables used as predictors were sea surface temperature ($^{\circ}$C), sea water salinity (psu), net primary production of biomass ($mg/m^3/day$), and dissolved molecular oxygen ($mmol/m^3$), obtained from the Copernicus Marine Environment Monitoring Service (\url{https://marine.copernicus.eu/}). We also included bathymetry (m) from the ETOPO 2022 Global Relief Model by NOAA. These variables were standardized to have mean 0 and standard deviation 1. We used the GLOSSA application to predict both the native range and the suitable habitat of \textit{Siganus luridus} in the year 2020. While the suitable habitat model is based only on environmental variables, the native range model also includes latitude and longitude as predictors.

\begin{figure}[H]
\centering
\includegraphics[width=\linewidth]{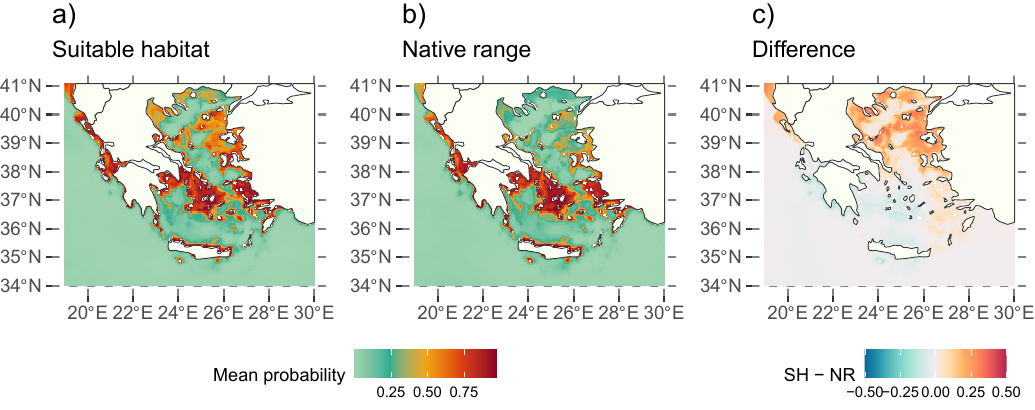}
\caption{Mean posterior probability for the suitable habitat and native range of \textit{Siganus luridus} in the Greek seas in 2020 and the difference between both predictions (SH: suitable habitat, NR: native range).}
\label{fig:siganus_luridus}
\end{figure}

Figure \ref{fig:siganus_luridus} shows the mean probability of the posterior predictive distribution for the native range and the suitable habitat, revealing differences between both models. The spatially smoothed native range predicts where the species is currently present, while the suitable habitat identifies areas with environmental conditions favorable for the species, even where it has not yet been recorded. For example, the Thracian Sea and northeast Aegean islands were identified as high-risk areas for potential invasion, even though the species is not currently established there according to the predicted native range in 2020. However, there's high uncertainty in the Thracian Sea predictions as the environmental conditions are outside the range of the ones matching the occurrence records \citep{solanou2023looking}. More information on this case study can be found in Supplementary File S3.

\subsection*{Memory usage and computation time}

The runtime and memory usage of GLOSSA are influenced by several factors that increase the complexity of the analysis, including the number of species analyzed in a single run, the number of species occurrences, the number of predictor variables, the size and resolution of the study area, and the number of requested projections. To provide some benchmarks, we report the computation time from the analysis of the worked examples. The analyses were conducted on a single Windows 11 machine equipped with 64 GB of RAM and an Intel(R) Core(TM) i7-1165G7 processor. This processor features 4 cores and 8 threads, with a base clock speed of 2.80 GHz. The global analysis of \textit{Thunnus albacares} took 9 hours and 9 minutes, while the regional models for \textit{Caretta caretta} and \textit{Siganus luridus} took 38 and 4 minutes, respectively. Detailed computation times for each step of the analysis are provided in the Supplementary Files S1–S3.

\section*{Conclusions and future directions}

Here, we present the \texttt{GLOSSA} R shiny app as a user-friendly tool for species distribution modeling. This open-source platform is a strong candidate for a broad audience within and outside the scientific community due to its intuitive app. GLOSSA stands out not only because of its simple and attractive interface but also because it includes innovative methodologies for species distribution modeling. Additionally, the application is well-documented with example tutorials and videos. One of the main strengths of GLOSSA is the BART modeling approach, which provides a flexible method to incorporate complex relationships between species occurrences and environmental conditions. It takes advantage of the Bayesian framework to regularize the fit and comprehensively explore the results of the analysis.

The examples presented here demonstrate the capability of GLOSSA to address complex ecological scenarios, such as evaluating the effects of climate change on species distribution shifts on a global scale or identifying areas with a higher potential risk of species invasion with a local focus. We analyzed how climate change may alter the global distribution of \textit{Thunnus albacares}, as well as the distribution of \textit{Caretta caretta} on a regional scale, focusing on the Mediterranean Sea, while also determining the most influential environmental variables. Additionally, GLOSSA was employed to identify potential invasion risk areas for \textit{Siganus luridus} in the Greek Seas. 

Although GLOSSA is currently designed for species distribution modeling, it represents an opportunity to be integrated into broader analysis workflows. For example, Marine Ecosystem Models (MEMs) are mechanistic (hybrid) models based on food webs and other ecosystem components that aim to study the environmental and anthropogenic factors that affect ecosystem responses \citep{geary2020guide, tittensor2018protocol, nader2022ecosystem}. MEMs rely on extensive parametrization and require a large amount of data to operate effectively \citep{robson2018towards, steenbeek2021making}. SDMs, such as those fitted by GLOSSA, can be particularly useful in this context, as they provide information on potentially suitable habitats and the functional relationships of species with their environments, which can be used to inform MEMs \citep{coll2019predicting, coll2020advancing}. Future work should explore this link.

When using species distribution models (SDMs), it is important to consider the limitations of these models, such as biases in occurrence sampling methods, issues with spatial autocorrelation between occurrences, the resolution and scale of the environmental layers, and temporal biases between the recorded occurrences and the environmental values assigned, among others.

Future work for GLOSSA will focus on improving
its functionality and ease of use by connecting to public databases, expanding data processing options and modeling approaches, including ensemble predictions, and modeling of additional data types. We will also aim to expand applications to other habitats, such as terrestrial and freshwater. Finally, documentation updates, as well as tutorials, latest changes, and improvements, will be available on the GLOSSA documentation webpage \url{https://imares-group.github.io/glossa/}.

\section*{Acknowledgements}

We would like to thank the iMARES group for their helpful feedback on the app design. This study is part of the ProOceans project (PID2020-118097RB-I00) funded by the Spanish Ministry of Science and Innovation. The authors also acknowledge the institutional support of the "Severo Ochoa Center of Excellence" accreditation (CEX2019-000928-S) to the Institute of Marine Science (ICM-CSIC). Additionally, this research is part of the Integrated Marine Ecosystem Assessments (iMARES) research group, funded by the Agència de Gestió d'Ajuts Universitaris i de Recerca of the Generalitat de Catalunya (2021 SGR 00435).

\section*{Conflict of Interest statement}

The authors declare that they have no conflict of interest.

\section*{Author Contributions}

J.M.T., A.F.A., J.M.B., and M.C. conceptualized and designed the GLOSSA application and the manuscript. J.M.T. developed and implemented GLOSSA, performed the analyses, and generated the visualizations. A.F.A. contributed to the GLOSSA workflow, particularly in the modeling approach using BARTs. J.M.B. and M.C. provided supervision, offering insights and guidance throughout the project. J.M.T. led the writing of the manuscript. All authors contributed critically to the drafts and approved the final version.

\section*{Data Availability}

The R package \texttt{glossa} is available on GitHub (\url{https://github.com/iMARES-group/glossa}) and the Comprehensive R Archive Network (CRAN; \url{https://cran.r-project.org/package=glossa}). The documentation for GLOSSA is accessible at \url{https://imares-group.github.io/glossa/}. Data and code used in the examples are available on Zenodo (\url{https://doi.org/10.5281/zenodo.13833556}). Occurrence data for \textit{Thunnus albacares} is available from the OBIS database (\url{https://obis.org/taxon/127027}), for \textit{Caretta caretta} from GBIF (\url{https://doi.org/10.15468/dl.es7562}), and for \textit{Siganus luridus} from the GreekMarineICAS geodataset (\url{https://doi.org/10.25607/t2smha}), created as part of the ALAS (Aliens in the Aegean – A Sea Under Siege) project.

\section*{Supporting Information}

Additional supporting information is provided for the worked examples:

\textbf{Supplementary file S1.} Worldwide suitable habitat of \textit{Thunnus albacares}.

\textbf{Supplementary file S2.} Distribution of the loggerhead sea turtle in the Mediterranean Sea.

\textbf{Supplementary file S3.} Potential risk areas of \textit{Siganus luridus} in the Greek Seas.

\bibliography{GLOSSA_preprint}

\end{document}